\documentclass{article}
\usepackage{spconf,amsmath,graphicx,hyperref}
\usepackage{cite}
\usepackage{booktabs}
\usepackage{amsmath}
\usepackage{tikz}
\usepackage{amsfonts}
\usepackage{amssymb}
\usepackage{dsfont} 
\usepackage{pifont}
\usetikzlibrary{arrows.meta,positioning}
\usepackage[table]{xcolor}

\title{EMO-TTA: Improving Test-Time Adaptation of Audio-Language Models for Speech Emotion Recognition}
\name{Jiacheng Shi$^{\star}$ \qquad Hongfei Du$^{\star}$ \qquad Y. Alicia Hong$^{\ddagger
}$ \qquad Ye Gao$^{\star}$}
\address{$^{\star}$ College of William \& Mary, $^{\ddagger}$ George Mason University\\
\small \texttt{\{jshi12, hdu02, ygao18\}@wm.edu, yhong22@gmu.edu}}

\begin{document}
\ninept
\maketitle
\begin{abstract}
Speech emotion recognition (SER) with audio-language models (ALMs) remains vulnerable to distribution shifts at test time, leading to performance degradation in out-of-domain scenarios. Test-time adaptation (TTA) provides a promising solution but often relies on gradient-based updates or prompt tuning, limiting flexibility and practicality. We propose Emo-TTA, a lightweight, training-free adaptation framework that incrementally updates class-conditional statistics via an Expectation-Maximization procedure for explicit test-time distribution estimation, using ALM predictions as priors. Emo-TTA operates on individual test samples without modifying model weights. Experiments on six out-of-domain SER benchmarks show consistent accuracy improvements over prior TTA baselines, demonstrating the effectiveness of statistical adaptation in aligning model predictions with evolving test distributions.
\end{abstract}

\section{Introduction}
\label{sec:intro}
Speech emotion recognition (SER)~\cite{ren2024emo, lin2025improving} serves as a foundational component in applications such as conversational agents, contact center analytics, healthcare, and education. However, real-world deployment often faces significant distributional shifts due to variations in speakers, languages, devices, ambient noise, or even inconsistent label taxonomies. As a result, models that perform well in-domain often degrade substantially on out-of-domain (OOD) test scenarios~\cite{bukhari2024selm}.

Test-time adaptation (TTA)~\cite{abdul2023align} adapts models during inference using only unlabeled target data, without source access. Early methods tune class-specific prompts with a few target labels for alignment~\cite{zhou2022learning,zhou2022conditional}. Later approaches remove labels but rely on gradient-based updates to learn prompts or adapters from unlabeled test batches~\cite{shu2022test,liang2023adapting}. Recent training-free methods eliminate all optimization by using heuristics or retrieval to improve robustness~\cite{zanella2024mtm,farina2024frustratinglyzero}. However, these methods still assume access to a batch of samples and introduce computational overhead during inference. While effective, most require batch-mode access and tuning at inference, limiting their use in practical applications.

While test-time adaptation has recently gained traction in speech, particularly for ASR~\cite{deshmukh2024domain,lin2022listen}, via unsupervised objectives, pseudo-labeling, and low-rank parameter tuning, its extension to SER remains relatively limited. Compared to ASR, SER exhibits greater speaker variability, emotional ambiguity, and weaker linguistic structure, posing unique adaptation challenges. Compounding this, many existing TTA methods rely on test batches~\cite{deshmukh2024domain}, model updates~\cite{lin2022listen}, or prompt tuning~\cite{zhou2022learning}, making them unsuitable when access to training sources or buffered inputs is severely restricted. This highlights the growing need for lightweight, training-free adaptation approaches that avoid parameter updates.


To address these challenges, we propose \emph{Emo-TTA}, a training-free test-time adaptation method that integrates audio language models (ALMs)~\cite{wu2023large,elizalde2023clap} with a test time Expectation-Maximization (EM)~\cite{moon1996expectation} procedure. Emo-TTA is designed to meet three key requirements: (i) \emph{Test-time distribution estimation}, by maintaining class-conditional statistics over the incoming test samples to inform predictions; (ii) \emph{Lightweight adaptation}, by avoiding access to source data, model updates, or multi-sample buffering; and (iii) \emph{Training-free inference}, without prompt tuning or retraining. Specifically, we model each class-conditional density \( p(x \mid y = i) \sim \mathcal{N}(\mu_i, \Sigma_i) \) with prior \( \pi_i \), and predict using Bayes rule: \( \hat{y} = \arg\max_i\, p(x \mid y=i)\, p(y=i) \). Initialization sets \( \mu_i \) from ALM-derived text prompts and uses simple priors \( \pi_i \). At each time step \( t \), the E-step computes soft assignments \( \gamma_{t,i} \propto \pi_i\, \mathcal{N}(x_t \mid \mu_i, \Sigma_i) \), and the M-step incrementally updates \( \{ \mu_i, \Sigma_i, \pi_i \} \) from \( \{ \gamma_{t,i}, x_t \} \). This yields continuous, single-sample adaptation without modifying the model or storing past test data.

In summary, our main contributions are:  
(i) To the best of our knowledge, we are the first to simultaneously satisfy the three core requirements of TTA in ALMs, namely Test-time distribution estimation, lightweight adaptation, and training-free inference, thereby enabling efficient and practical adaptation for SER. 
(ii) We introduce \emph{Emo-TTA}, a EM-based 
 method that incrementally updates class-conditional statistics based on ALM-derived textual prototypes, enabling continuous adaptation without accessing source data, modifying model weights, or storing test samples.  
(iii) Extensive experiments on six out-of-domain SER benchmarks show that Emo-TTA consistently improves performance, highlighting its robustness across acoustic domains.

\begin{figure*}[htbp]
    \centering
    \vspace{-2mm}
    \includegraphics[width=0.9\textwidth]{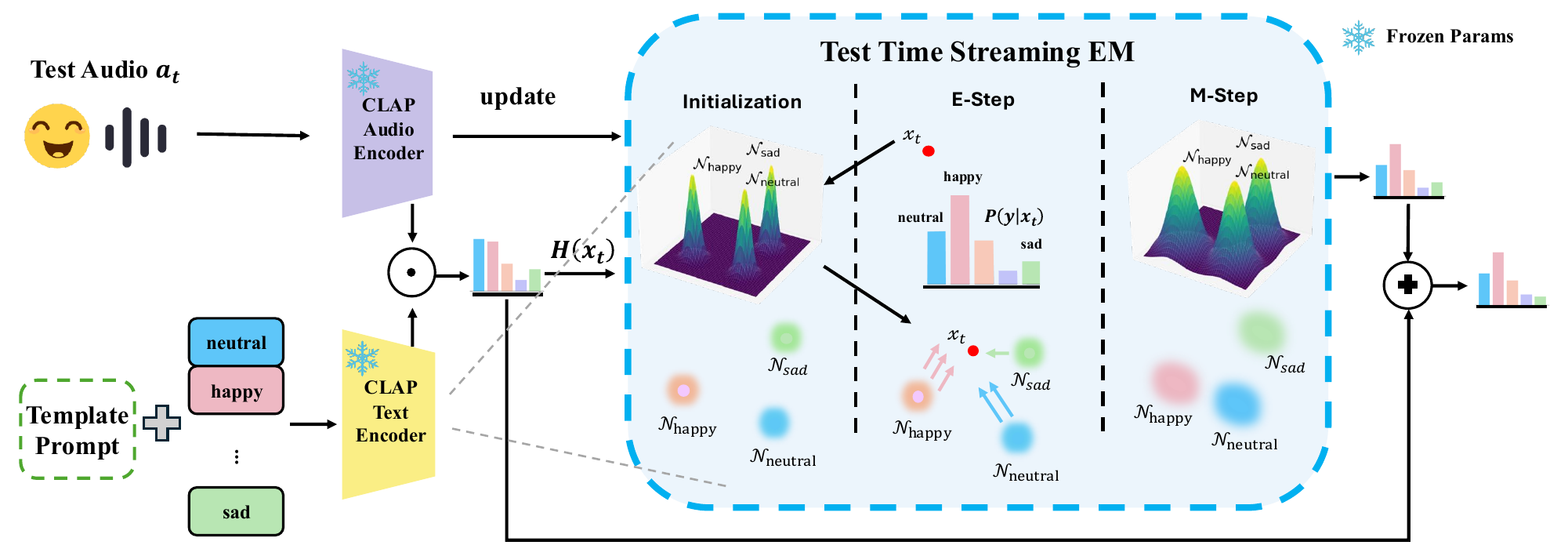}
    \vspace{-2mm}
    \caption{Overview of Emo-TTA for test-time adaptation in SER. Given a test audio, frozen CLAP encoders extract audio and class prototypes, which initialize a EM-based that continuously updates Gaussian parameters via entropy-weighted confidence. Final prediction fuses CLAP’s zero-shot logits with generative scores for stable, training-free adaptation under distribution shift.}
    \label{fig:main}
    \vspace{-2mm}
\end{figure*}
\section{Method}
\subsection{Preliminaries}
Audio--Language Models (ALMs) such as CLAP achieve strong zero-shot performance by aligning acoustic and textual modalities. In SER, CLAP classifies utterances by computing the similarity between audio features and text prompts describing each emotion:
\begin{equation}
\small
P_{\text{CLAP}}(y=i \mid a)=
\frac{\exp\!\big(\cos\!\big(f(a),\, g(t_i)\big)\big)}
{\sum_{k=1}^{K}\exp\!\big(\cos\!\big(f(a),\, g(t_k)\big)\big)} ,
\end{equation}
where $f(\cdot)$ and $g(\cdot)$ are the audio and text encoders, and $\cos(\cdot,\cdot)$ denotes cosine similarity. $t_i$ is a class-specific text prompt for the $i$-th emotion, and $K$ is the number of emotion classes. Despite its strengths, CLAP exhibits performance degradation under acoustic domain shifts (e.g., speaker, channel, or noise). While prior methods address this via fine-tuning on labeled target-domain data, they incur high adaptation costs. In contrast, our training-free test-time approach enables lightweight adaptation, improving practical feasibility under real-world constraints.

\subsection{Test-Time Adaptation (TTA) for ALMs.}
In the TTA setting, an audio--language model pre-trained on a source domain is adapted to a target \emph{speech emotion recognition} (SER) domain using only unlabeled test audio $D_{\text{test}}=\{a_t\}$, with predictions made independently at inference. To improve CLAP’s performance under distribution shift, prior TTA methods aggregate predictions over multiple \emph{audio augmentations} via entropy-based filtering:
\begin{equation}
\small
\begin{aligned}
&P^{*}(y \mid x_t) = \\
&\frac{1}{\rho M} \sum_{m=1}^{M} 
   \mathds{1}\!\left[\mathcal{H}\!\left(P_{\text{CLAP}}(A_m(x_t))\right) \le \tau\right]\cdot P_{\text{CLAP}}\!\left(A_m(x_t)\right)
\end{aligned}
\end{equation}
where $A_m(a_t)$ is the $m$-th augmented view of utterance $a_t$, $\rho$ is the selected proportion of low-entropy views, and $\mathcal{H}(\cdot)$ denotes entropy over $K$ emotion classes.

Despite promising results, existing approaches face limitations: (1) they neglect modeling the underlying emotion distribution, overlooking the structure among test utterances; (2) they rely on access to multiple samples or memory caches, incompatible with privacy-sensitive settings; and (3) they require test-time optimization, which introduces computational overhead and limits their applicability in settings without model updates or labeled data in SER task.

\subsection{Gaussian Discriminant Analysis}

Assume CLAP audio embeddings for each emotion class $y$ follow a multivariate normal distribution with class-specific means and shared covariance. Let $F_t = f(a_t) \in \mathbb{R}^d$ be the CLAP \emph{audio} embedding for the $t$-th utterance, with $\mu_y$ denoting the mean embedding of class $y$, and $\Sigma$ the shared covariance matrix. The class-conditional likelihood is
{\small
\begin{align}
& p(F_t \mid y) = \mathcal{N}(F_t \mid \mu_y, \Sigma) \nonumber \\
&= \frac{1}{(2\pi)^{d/2} |\Sigma|^{1/2}}
   \exp\!\left( -\frac{(F_t - \mu_y)^\top \Sigma^{-1} (F_t - \mu_y)}{2} \right),
\label{cdf}
\end{align}
}
where $d$ is the embedding dimension. This likelihood captures the Mahalanobis distance between $F_t$ and $\mu_y$, adjusted for feature correlation and scale. Applying Bayes’ rule yields the posterior:
\begin{equation}
\small
P(y \mid F_t) \;=\; \frac{P(y)\,p(F_t \mid y)}{\sum_{y'} P(y')\,p(F_t \mid y')},
\tag{4}
\end{equation}
and substituting Eq.~\eqref{cdf} leads to the softmax form:
\begin{equation}
\small
P(y \mid F_t) \;=\;
\frac{\exp\!\left(-\tfrac{1}{2}(F_t-\mu_y)^\top \Sigma^{-1} (F_t-\mu_y)\right)}
{\sum_{y'} \exp\!\left(-\tfrac{1}{2}(F_t-\mu_{y'})^\top \Sigma^{-1} (F_t-\mu_{y'})\right)}.
\tag{5}
\end{equation}
For the $t$-th test utterance sample, the final prediction corresponds to the class with highest posterior probability:
\begin{equation}
\small
y^* \;=\; \arg\max_{y} \left( \log P(y) \;-\; \tfrac{1}{2}\,(F_t-\mu_y)^\top \Sigma^{-1} (F_t-\mu_y) \right).
\tag{6}
\end{equation}

\subsection{EM-based Algorithm for TTA}
Although GDA~\cite{bishop2006pattern} captures the test time distribution, it typically requires labeled data for reliable parameter estimation. In TTA for SER, however, test instances are unlabeled and arrive sequentially, making direct use of GDA impractical. To address this, we introduce a EM-based algorithm leveraging CLAP’s zero-shot predictions as soft priors to iteratively estimate posteriors and update parameters on-the-fly.
\subsubsection{Parameter Initialization}
At initialization, the CLAP text encoder $g(\cdot)$ generates a semantic prototype $g(t_y)$ for each emotion class $y$, which we adopt as the initial class mean, i.e., $\mu_y = g(t_y)$. Assuming the embeddings are i.i.d.\ with unit variance, we simply set the shared covariance matrix to the identity, i.e., $\Sigma = I$.

\subsubsection{E-Step: Posterior Inference}
Given the CLAP \emph{audio} embedding $F_t = f(a_t)$ for the $t$-th test sample, we compute its posterior responsibility $\gamma_{y,t} = P(z_y = 1 \mid F_t)$ under the Gaussian assumption. Specifically,
\begin{equation}
\small
P_{\text{GAUS}}(z_y=1 \mid F_t)
= \frac{\pi_y \, \mathcal{N}\!\big(F_t \mid \mu_y, \Sigma\big)}
       {\sum_{j} \pi_j \, \mathcal{N}\!\big(F_t \mid \mu_j, \Sigma\big)} \,,
\tag{8}
\end{equation}
where $z_y$ indicates class membership and $\pi_y$ is the class prior.

\subsubsection{M-Step: Parameter Updates}
In the M-step, we update the class prior $\pi_y$, class mean $\mu_y$, and shared covariance $\Sigma$ in the CLAP embedding space based on the posterior responsibility $\gamma_{y,t}$ of test instance $t$. Let $F_t = f(a_t)$ denote the CLAP audio embedding of the $t$-th test sample. Denote $n_t$ as the cumulative number of test samples observed so far, and initialize the effective class count $N_y$ to $1/K$. All parameter updates are computed jointly as follows:
\begin{equation}\tag{9}
\small
\begin{gathered}
\pi_y’ = \frac{N_y + \gamma_{y,t}}{n_t}, \\
\mu_y’ = \frac{N_y,\mu_y + \gamma_{y,t},F_t}{N_y + \gamma_{y,t}}, \\
\Sigma’ = \frac{(n_t{-}1)\Sigma + \sum_{y} \gamma_{y,t}(F_t{-}\mu_y’)(F_t{-}\mu_y’)^\top}{n_t{-}1}
\end{gathered}
\end{equation}
We update the effective count as $N_y' = N_y + \gamma_{y,t}$. This EM-based procedure adapts class statistics incrementally without supervision or gradient descent, enabling efficient TTA under distribution shift.

\subsection{Incorporating ALM Priors}
To stabilize early-stage TTA for SER, we incorporate CLAP priors by integrating zero-shot predictions and their confidence into the update process. Early predictions may deviate from emotion semantics due to distribution shift. To mitigate this, we estimate each utterance’s uncertainty via the entropy of CLAP’s predicted distribution and modulate its influence accordingly. Specifically, for the $t$-th test utterance $a_t$, we compute its self-entropy from CLAP’s predicted probabilities ${,P_{\text{CLAP}}(z_y=1 \mid a_t)\mid y=1,\ldots,K,}$. The entropy is \begin{equation}\tag{10}
\small
H(a_t) \;=\; -\sum_{y=1}^{K} P_{\text{CLAP}}(z_y=1 \mid a_t)\,\log P_{\text{CLAP}}(z_y=1 \mid a_t).
\end{equation}
We define the confidence-based weight $w(h)=e^{-\beta h}$ (with $\beta>0$) and revise the EM updates as:
\begin{equation}\tag{11}
\small
\begin{gathered}
\pi_y' \;=\; \dfrac{N_y + w\!\big(H(a_t)\big)\cdot \gamma_{y,t}}{\,n_t' + w\!\big(H(a_t)\big)\,},\\[4pt]
\mu_y' \;=\; \dfrac{N_y\cdot \mu_y + w\!\big(H(a_t)\big)\cdot \gamma_{y,t}\cdot F_t}{\,N_y + w\!\big(H(a_t)\big)\cdot \gamma_{y,t}\,},\\[4pt]
\Sigma' \;=\; \dfrac{(n_t'-1)\Sigma \;+\; w\!\big(H(a_t)\big)\,\displaystyle\sum_{y=1}^{K}\gamma_{y,t}\big(F_t-\mu_y'\big)\big(F_t-\mu_y'\big)^{\!\top}}{\,n_t'-1\,},
\end{gathered}
\end{equation}
where $F_t=f(a_t)$ is the CLAP audio embedding and $\gamma_{y,t}$ is the posterior responsibility. After incorporating $a_t$, the counts are updated as
$N_y' = N_y + w\!\big(H(a_t)\big)\cdot \gamma_{y,t}$ and $n_t' = n_{t-1}' + w\!\big(H(a_t)\big)$. This entropy-aware mechanism reduces the impact of uncertain utterances and mitigates noise-induced drift. Finally, we combine CLAP’s zero-shot logits with those from the generative model:
\begin{equation}\tag{12}
\small
\text{logits}_y \;=\; T_y^{\top}F \;+\; \alpha\big(w_y^{\top}F + b_y\big),
\end{equation}
where $F=f(a_t)$ is the audio feature, $T_y=g(t_y)$ is the text prototype, $\alpha$ is a tunable coefficient, and $w_y=\Sigma^{-1}\mu_y$, $b_y=\log P(y)-\tfrac{1}{2}\mu_y^{\top}\Sigma^{-1}\mu_y$ from the generative model.

\section{Experiments}
\renewcommand\arraystretch{0.95}
\begin{table*}[ht]
  \centering
  \resizebox{0.8\linewidth}{!}{
    \begin{tabular}{l*{9}{c}c}
      \toprule
      \textbf{Method} & T.F. & L.W. & Est. & IEMOCAP & MELD & RAVDESS & TESS & SAVEE & CREMA\text{-}D & AVG \\
      \midrule
      CLAP\text{-}PANN-14~\cite{wu2023large} & - & - & - & 34.52 & 17.11 &18.91 & 49.76 & 38.38 & 29.54 & 31.37 \\
      \midrule
      CoOp~\cite{zhou2022learning}   & \ding{55} & \ding{55} & \ding{55} & 33.83 & 16.47 & 19.03 & 49.73 & 37.66 & 33.55 & 31.71 \\
      CoCoOp~\cite{zhou2022conditional} & \ding{55} & \ding{55} & \ding{55} & 33.72 & 17.28 & 22.57 & 50.71 & 38.93 & 36.36 & 33.26 \\
      Treff-Adapter~\cite{liang2023adapting} & \ding{55} & \ding{55} & \ding{55} & 35.86 & 18.85 & 26.45 & 52.84 & 42.03 & \textbf{40.59} & 36.11 \\
      TPT~\cite{shu2022test}     & \ding{55} & \ding{51} & \ding{55} & 35.58 & 17.77 & 25.53 & 50.34 & 39.95 & 35.03 & 34.03 \\
      
      \midrule
      \textbf{Ours} & \ding{51} & \ding{51} & \ding{51} & \textbf{39.92} & \textbf{19.91} & \textbf{29.54} & \textbf{54.54} & \textbf{44.76} & 39.44 & \textbf{38.02} \\
      \midrule
      \midrule
      CLAP\text{-}HTSAT~\cite{wu2023large} & - & - & - & 36.35 & 18.62 & 19.86 & 50.31 & 39.25 & 31.07 & 32.57 \\
      \midrule
      CoOp~\cite{zhou2022learning}   & \ding{55}   & \ding{55} & \ding{55} & 34.77 & 17.73 & 19.14 & 50.63 & 37.67 & 33.12 & 32.17 \\
      CoCoOp~\cite{zhou2022conditional} & \ding{55} & \ding{55} & \ding{55} & 35.42 & 18.57 & 23.93 & 50.92 & 39.13 & 36.74 & 34.18 \\
      Treff-Adapter~\cite{liang2023adapting} & \ding{55} & \ding{55} & \ding{55} & 38.13 & 19.61 & 27.04 & 53.91 & 42.32 & 41.61 & 37.10 \\
      TPT~\cite{shu2022test}     & \ding{55} & \ding{51} & \ding{55} & 36.80 & 18.89 & 26.63 & 50.58 & 40.76 & 36.05 & 34.96 \\
      MTA~\cite{zanella2024mtm}     & \ding{51} & \ding{51} & \ding{55} & 38.92 & 18.93 & 25.74 & \textbf{56.75} & 40.96 & 34.94 & 36.04 \\
      ZERO~\cite{farina2024frustratinglyzero} & \ding{51} & \ding{51} & \ding{55} & 37.85 & 18.50 & 26.02 & 55.47 & 41.44 & 34.88 & 35.69 \\
      \midrule
      \textbf{Ours} & \ding{51} & \ding{51} & \ding{51} & \textbf{43.65} & \textbf{20.17} & \textbf{31.72} & 56.09 & \textbf{46.39} & \textbf{44.78} & \textbf{40.47} \\
      \bottomrule
    \end{tabular}
    }
  \caption{Comparison of out-of-domain SER accuracy (\%) across datasets. Methods are grouped by training-free (T.F.), lightweight adaptation (L.W.), and test-stream distribution estimation (Est.). Bold indicates the highest accuracy per dataset.}
  \label{tab:ser_cross_domain}
  \vspace{-1pt}
\end{table*}





\begin{table}[t]
\centering
\scriptsize
\begin{tabular}{l c c c c}
\toprule
\textbf{Model} & \textbf{IEMOC} & \textbf{CREMA} & \textbf{RAVD} & \textbf{AVG} \\
\midrule
CLAP~\cite{wu2023large}   & 36.35 & 31.07 & 19.86 & 29.09 \\
Pengi~\cite{deshmukh2023pengi}   & 35.63 & 33.46 & 23.07 & 30.72 \\
MMS large~\cite{ma2023investigating_wisper}   & 36.89 & 32.20 & 19.33 & 29.47 \\
Whisper medium.en~\cite{ma2023investigating_wisper}   & 37.90 & 34.60 & 19.02 & 30.51 \\
Whisper medium~\cite{ma2023investigating_wisper}   & 36.81 & 34.09 & 20.14 & 30.35 \\
Whisper large-v2~\cite{ma2023investigating_wisper}   & 38.10 & 35.80 & 19.59 & 31.16 \\
AudioFlamingo~\cite{kong2024audio_flamingo}   & - & - & 23.44 & - \\
EmoCLAP~\cite{dhamyal2022describing_emoclap}   & - & 35.22  & -  & - \\
SELM~\cite{bukhari2024selm}   & 40.02 & 42.79 & 24.51  & 35.77 \\
\midrule
\textbf{Emo-TTA} & \textbf{43.65} & \textbf{44.78} & \textbf{31.72} & \textbf{40.05} \\

\bottomrule
\end{tabular}
\caption{Out-of-domain performance of different models across three datasets, where the target dataset is fully excluded from both training and any form of unsupervised test-time adaptation.}
\vspace{-4mm}
\label{tbl:out-of-domain-ALM}
\end{table}
\subsection{Datasets and Implementation Details}
\textbf{Datasets. }We study speech emotion recognition (SER) in a cross-corpus out-of-domain (OOD) setting.
IEMOCAP~\cite{busso2008iemocap}, a dyadic scripted/improvised dialogue dataset with 10 speakers, follows a 5-fold leave-one-session-out cross-validation setup, merging excited into happy where applicable with a unified 4-class label space (angry, happy, sad, neutral) following~\cite{shi2025clep}.
MELD~\cite{poria2018meld}, derived from multi-party dialogues in Friends, is evaluated using its official train/dev/test split, with emotions mapped to the unified 4-class scheme.
RAVDESS~\cite{livingstone2018ryersonRAVDESS} contains 2.5k studio-quality utterances  averaging 5s, and we retain all original 8 emotion classes using a 5-fold stratified cross-validation protocol.
TESS~\cite{dupuis2010toronto} and SAVEE~\cite{SAVEE}, both acted emotion datasets with consistent recording conditions, are evaluated using random 10-fold cross-validation with 4-class mapping. CREMA-D~\cite{cao2014crema}, containing 7k utterances averaging 4.5s, uses a 5-fold CV setup and retains its original 6 emotion classes, following\cite{bukhari2024selm}.
This diverse selection enables a rigorous evaluation of model generalization across varying speaker identities, acoustic conditions, and emotional taxonomies.

\noindent\textbf{Implementation Details.}We resample all audio to 16~kHz and standardize input length to 5 seconds via truncation or zero-padding. Each clip is paired with a text label in the form ``This is a [EMOTION] sound'' to ensure cross-modal alignment. We use CLAP with PANN-14 or HTS-AT as the audio encoder and RoBERTa as the text encoder. CoOp, CoCoOp, and Treff Adapter are implemented following prior work~\cite{li2024audio,liang2023adapting}, while TPT~\cite{shu2022test}, MTA~\cite{zanella2024mtm}, and ZERO~\cite{farina2024frustratinglyzero} are adapted by replacing original encoders with CLAP encoders following by~\cite{elizalde2023clap}. For Emo-TTA, we set $\alpha{=}0.2$, $\beta{=}4.5$, and follow a test time emotional TTA setup with batch size 1 and no backpropagation. We report the mean top-1 classification accuracy, averaged over three runs with random seeds.
\subsection{Main Results}
\noindent\textbf{Comparison with prior TTA methods.}
Table~\ref{tab:ser_cross_domain} reports out-of-domain SER performance across six datasets under two CLAP backbones: PANN-14~\cite{kong2020panns} and HTS-AT~\cite{chen2022hts}. CoOp~\cite{zhou2022learning} and CoCoOp~\cite{zhou2022conditional} are few-shot prompt-learning methods requiring labeled target data. Treff-Adapter~\cite{liang2023adapting} and TPT~\cite{shu2022test} perform gradient-based test-time tuning, while MTA~\cite{zanella2024mtm} and ZERO~\cite{farina2024frustratinglyzero} are training-free, source-free baselines that do not explicitly model the test distribution. In contrast, Emo-TTA conducts sample-wise, optimization-free adaptation through a EM-based procedure that explicitly estimates the test-time distribution. Under CLAP-PANN-14, Emo-TTA achieves the best average accuracy of 38.02\%, outperforming Treff-Adapter (36.11) by +1.91 and the zero-shot CLAP baseline (31.37) by +6.65, and ranks first on 5 of 6 datasets (except CREMA-D). Under CLAP-HTSAT, Emo-TTA again leads with 40.47\%, surpassing Treff-Adapter (37.10), MTA (36.04), and ZERO (35.69), and achieves the top score on 5 of 6 datasets (TESS led by MTA). Overall, Emo-TTA delivers the best results on 10 out of 12 dataset/backbone combinations, consistently outperforming both prompt-learning and test-time adaptation baselines in the out-of-domain setting. These gains stem from its continuous EM procedure, which incrementally updates class statistics and calibrates predictions without training or prompt tuning, enabling robust test-time adaptation under distribution shifts.

\noindent\textbf{Comparison with Foundation Audio Models.}
We evaluate Emo-TTA under a strict out-of-domain setting against recent audio-language models. Pengi formulates audio understanding as prompt-based text generation using transfer learning. Whisper is a multilingual encoder-decoder trained on large-scale speech corpora. AudioFlamingo integrates audio into LLMs with in-context learning capabilities. SELM aligns audio and text inputs via a frozen language model for emotion prediction. Following prior work~\cite{bukhari2024selm}, we adopt a 4-class and 6-class emotion classification setup for IEMOCAP and CREMA-D, and an 8-class setup for RAVDESS. All target datasets are held out from training and adaptation. As shown in Table~\ref{tbl:out-of-domain-ALM}, Emo-TTA achieves the highest accuracy across all three datasets, with a top average of 40.05\%, outperforming the best baseline SELM (35.77\%) by +4.28. It yields substantial gains over the base CLAP model: +7.3\% on IEMOCAP, +13.71\% on CREMA-D, and +11.86\% on RAVDESS. These consistent improvements demonstrate the strength of our optimization-free EM-based procedure, which dynamically updates class statistics at test time without labeled data or gradient-based adaptation.

\begin{table}[t]
\centering
\scriptsize
\begin{tabular}{l c c c c}
\toprule
\textbf{Model} & \textbf{IEMOC} & \textbf{CREMA} & \textbf{RAVD} & \textbf{AVG} \\
\midrule
\textbf{Emo-TTA} & \textbf{43.65} & \textbf{44.78} & \textbf{31.72}  & \textbf{40.05} \\
Zero-shot CLAP~\cite{wu2023large}   & 36.35 & 31.07 & 19.86 & 29.09 \\
\midrule
\quad w/out Mean Vectors Update        & 42.33 & 43.37 & 29.51 & 38.40 \\
\quad w/out Covariance Matrix Update   & 37.94 & 37.39 & 25.58 & 33.64 \\
\quad w/out ALM priors   & 38.69 & 38.15 & 26.67 & 34.51 \\
\bottomrule
\end{tabular}
\caption{Ablation study demonstrates the impact of each Emo-TTA component on SER accuracy (\%).}
\vspace{-5mm}
\label{tbl:ablations}
\end{table}
\section{Ablation Study}
\noindent\textbf{Update of Class Means.}
To assess the role of adaptive class means $\mu_y$, we conduct an ablation by freezing $\mu_y$ to CLAP-derived text prototypes during evaluation. This static configuration underperforms under out-of-domain conditions, suggesting that fixed zero-shot anchors cannot track the shifting acoustic patterns of unseen target data. In contrast, dynamically updating $\mu_y$ over test instances allows the model to re-center class representations and better align with test-time distributions. This highlights the importance of adaptation through implicit structure discovery across individual test samples.

\noindent\textbf{Update of Covariance.}
To assess the impact of covariance adaptation, we conduct an ablation study where the covariance matrix $\Sigma$ is fixed as an identity matrix during evaluation (Tab.~\ref{tbl:ablations}). This simplification reduces the scoring function to Euclidean distance, constraining the model’s capacity to capture intra-class variability. As a result, the performance deteriorates under distribution shift. In contrast, updating $\Sigma$ dynamically allows the model to more accurately represent the dispersion and correlation patterns of class features, leading to improved test-time recognition accuracy.

\noindent\textbf{ALM Priors.}
To evaluate the influence of audio–language priors and confidence-aware updates, we initialize class prototypes with CLAP-derived text embeddings and modulate their updates using prediction confidence measured via self-entropy (Tab.~\ref{tbl:ablations}). Without prior-based initialization, the model suffers from poor alignment between prototypes and emotion classes, leading to performance degradation. In addition, disabling confidence weighting causes uncertain predictions to skew class statistics, reducing stability. Incorporating both components enables the model to adapt more robustly under distribution shifts by preserving semantic grounding and filtering noisy updates.

\section{Conclusion}
We introduce Emo-TTA, a training-free adaptation method that improves the robustness of ALMs for speech emotion recognition under distribution shifts. By modeling test-time distributions through lightweight EM updates and using ALM-based predictions as uncertainty-aware priors, Emo-TTA incrementally refines class statistics without accessing source data or updating model weights. Unlike prior TTA approaches, it operates on individual samples without prompt tuning or batched inputs. Experiments across six out-of-domain benchmarks confirm consistent performance gains, validating its effectiveness for practical emotion understanding.

\bibliographystyle{IEEEbib}
\bibliography{strings,main}

\end{document}